\documentclass[%
 aip,
 jmp,%
 amsmath,amssymb,
 reprint,%
]{revtex4-2}

\usepackage{graphicx}
\usepackage{dcolumn}
\usepackage{bm}
\usepackage{xcolor}
\usepackage{hyperref}

\usepackage{amsmath}
\usepackage{dutchcal}
\renewcommand{\d}{\text{d}}
\begin{document}

\preprint{AIP/123-QED}
\title[Waves in Bopp-Land{\'e}-Thomas-Podolsky generalized electrodynamics]{ Waves in Bopp-Land{\'e}-Thomas-Podolsky generalized electrodynamics}
\author{Altin Shala}
 \affiliation{ZARM, University of Bremen,
             28359 Bremen, Germany, {\tt altin.shala@zarm.uni-bremen.de}}
\author{Volker Perlick}%
\affiliation{Faculty 1, University of Bremen, 28359 Bremen, Germany, {\tt volker.perlick@uni-bremen.de}}%

\date{\today}
\begin{abstract}
\noindent We investigate the feasibility of probing Bopp-Land{\'e}-Thomas-Podolsky generalized electrodynamics with traveling and standing wave experiments. We consider wave propagation in vacuum and in a cold and non-magnetized plasma. Dispersion relations are found for all possible transverse and longitudinal modes. Longitudinal traveling waves are found which exhibit negative group velocities.
\end{abstract}

\maketitle
\section{Introduction}
\noindent 
Standard Maxwell electromagnetism is one of the most successful theories in physics. However, it shows various pathological features if one considers charged point particles as the sources of the electromagnetic field. All these pathologies have their origin in the fact that, according to the standard Maxwell theory, the field energy inside an arbitrary small sphere around a point charge is infinite. As a consequence, the self-force experienced by an electron comes out infinite if the latter is modeled as a classical pointlike particle. This has led Dirac [\onlinecite{dirac}] to assigning an infinite negative bare mass to the electron. However, the resulting equation of motion shows pathologies such as run-away solutions and pre-acceleration which have not been completely cured until now. One could avoid these problems by treating the electron as an extended body but this leads to equations of motion which are difficult to handle at relativistic velocities [\onlinecite{Yaghjian}]. For this reason it is desirable to model electrons as classical point particles, in particular in accelerator physics.

This has motivated several scientists to suggesting a modification of the standard Maxwell theory. Typically, these modified theories change the constitutive law of vacuum in such a way that the field energy inside a sphere around a point charge
becomes finite. The first attempt in this direction was made by Mie [\onlinecite{Mie:1913ice}] but his theory proved abortive. The best known modified theory of electromagnetism was suggested by Born and Infeld [\onlinecite{Born:1934gh}] as a direct successor and covariant generalization of Mie's original approach. They introduced a non-linear constitutive relation for the vacuum which contains a new hypothetical constant of Nature, $b$,  with the dimension of a magnetic field strength. For $b \to \infty$ the standard Maxwell constitutive law of the vacuum is recovered, while for finite $b$ the field of a point charge at rest in an inertial system is regularized in the sense that the energy in a sphere around such a charge becomes finite, as was demonstrated by Born and Infeld. Unfortunately, the  Born-Infeld theory is infamous for being very complicated because of its nonlinear constitutive relation. \\
In this paper we will be interested in another modified theory of electrodynamics that was originally introduced and discussed in great detail by Bopp [\onlinecite{bopp}]. It was rediscovered by Land{\'e} and Thomas [\onlinecite{PhysRev.60.121, PhysRev.60.514}] and then by Podolsky [\onlinecite{PhysRev.62.68}], so we call it the Bopp-Land{\'e}-Thomas-Podolsky (BLTP) theory. In this theory the constitutive law of the vacuum contains higher derivatives while retaining linearity. Similarly to the Born-Infeld theory, it introduces a new hypothetical constant of Nature, this time with the dimension of a length. This constant of Nature, denoted $l$, is properly called the \emph{Bopp length}. For $l \to 0$ the standard Maxwell theory is recovered while for $l \neq 0$ the field energy around a point charge is finite which leads to a finite self-force. This was shown for point charges at rest in an inertial system already in the original work by Bopp. For calculations of the self-force of accelerating point charges we refer to Zayats [\onlinecite{ZAYATS201411}] and to Gratus, Perlick and Tucker [\onlinecite{Gratus_2015}].\\
Theories with higher derivatives, such as the BLTP theory, are often criticized because the field energy is not bounded below (see e.g. Pais and Uhlenbeck [\onlinecite{pais_uhlen_1950}]) which leads to Ostrogradski instabilities. However, this is not a convincing argument as long as one views the BLTP theory as a classical effective theory (that could be derived as the classical limit of a yet-to-be-found modified quantum electrodynamics) rather than as a fundamental theory. It is well known that metamaterials exist which are well described by  effective field theories with field energies that are unbounded below. \\
The fact that to date no deviations from the standard Maxwell theory have been observed implies that the Bopp length must be small. Specific bounds on $l$ have been found with various methods. One of these methods consists in probing deviations from Coulomb's law by means of an ion interferometer, which was originally suggested by Neyenhuis, Christensen and Durfee [\onlinecite{PhysRevLett.99.200401}] for testing the Proca theory. Cuzinatto et al. [\onlinecite{CUZINATTO_2011}] applied this method to the BLTP theory and came to the conclusion that the accuracy is not sufficient for finding competitive bounds on the Bopp length. In the same paper Cuzinatto et al. also used the Rayleigh-Ritz method for calculating the modification of the ground state of the hydrogen atom according to the BLTP theory. A more systematic study of the BLTP hydrogen spectrum was carried through by Carley, Kiessling and Perlick [\onlinecite{doi:10.1142/S0217751X1950146X}]. It was shown that the corresponding modifications of the hydrogen spectrum would have been observed already if $l$ were bigger than $10^{-18} \, \mathrm{m}$. To the best of our knowledge, this is the smallest bound on $l$ we have to date. \\
The method of mirror charges and the Casimir effect in the BLTP theory has been studied by Barone and Nogueira [\onlinecite{Barone:2016pyy}]. By probing the Stefan-Boltzman law another constraint on $l$ has been found by Bonin et. al. [\onlinecite{Bonin_etal}]. For more context in terms of quantum physics we refer to a paper of Ji et al. [\onlinecite{Ji_2019}] which compares classical BLTP theory with the Pauli-Villars regularization procedure used in Maxwell quantum electrodynamics [\onlinecite{pauli}].\\
In this paper we want to study electromagnetic waves in flat spacetime according to the BLTP theory. Some basic results in this direction have been found already in the original paper by Bopp, but here we present a complete discussion of transverse and longitudinal modes, of traveling and standing waves, not only in vacuum but also in the presence of a cold and non-magnetized plasma. In this context it is interesting to note that vacuum waves in the BLTP theory are formally similar to waves in a plasma according to the standard Maxwell theory, with the Bopp length $l$ corresponding to a constant plasma frequency $\omega _p = c/l$. This analogy was observed already by Santos [\onlinecite{SANTOS_2011}]. In this paper we will study the combined effect of a real plasma and the Bopp length that is analogous to a plasma. We also refer to Elskens and Kiessling [\onlinecite{Elskens_2020}] where the respective kinetic plasma theory was studied in terms of its microscopic foundations.\\
Our study of wave propagation in vacuum and in a medium according to the BLTP theory should also be compared with analogous investigations in the Born-Infeld theory. We refer, in particular to Ferraro [\onlinecite{Ferraro_2007}] who considered waveguides and to Manojlovic, Perlick and Potting [\onlinecite{Manojlovic_2020}] who considered standing waves in the Born-Infeld theory. Fluids in the context of generalized electrodynamics have been studied by Dereli and Tucker [\onlinecite{Dereli_2010}] for the Born-Infeld theory.

\section{The BLTP field equations}
\noindent
In standard vector notation, the generic Maxwell equations on Minkowski spacetime read, in Heaviside-Lorentz units,
\begin{align}
    &\nabla \cdot \vec{B} = 0 \, ,
\label{eq:Max3}
\\
    &\nabla \times \vec{E} + \frac{1}{c}\frac{\partial \vec{B}}{\partial t} = 0 \, .
\label{eq:Max4}
\\
    &\nabla \cdot \vec{D} =   \rho \, , 
\label{eq:Max1}
\\
    &\nabla \times \vec{H} - \frac{1}{c}\frac{\partial \vec{D}}{\partial t} = \frac{1}{c} \vec{j}  \, ,
    \label{eq:Max2}
\end{align}
Here $\vec{E}$ and $\vec{B}$ are the field strengths, $\vec{D}$ and $\vec{H}$ are the excitations, $\rho$ is the charge density and $\vec{j}$ is the current density. These equations have to be supplemented with constitutive relations which specify the medium. In this paper we consider the constitutive relations for vacuum, but not in the standard Maxwell theory but rather in the BLTP theory, where the constitutive relations are given by
\begin{align}
    &\vec{D} = \vec{E} - l^2 \Box \vec{E} 
\label{eq:constrel1}
\\
    &\vec{H} = \vec{B} - l^2 \Box \vec{B} \, ,
\label{eq:constrel2}
\end{align}
Here $\Box$ denotes the d'Alembert operator on Minkowski spacetime,
\begin{equation}
    \Box     = -\frac{1}{c^2}\partial_t^2 + \Delta \, ,  \quad
    \Delta =  \partial_x^2 + \partial_y^2 + \partial_z^2 \, ,
\end{equation}
and $l$ is a new hypothetical constant of Nature with the dimension of a length; it is properly called the \emph{Bopp length}. We observe that Eqs. (\ref{eq:constrel1}) and (\ref{eq:constrel2}) retain linearity but introduce higher derivative terms: If inserted into Eqs. (\ref{eq:Max1}) and (\ref{eq:Max2}), the resulting equations involve third-order derivatives of the field strengths.

In analogy to the standard Maxwell theory, the Maxwell equations with the BLTP constitutive relations imply wave equations for $\vec{E}$ and for $\vec{B}$ alone. We begin by inserting the constitutive relations into Eq. (\ref{eq:Max2}) and applying another time derivative, 
\begin{align}
    &\nabla \times\left( \frac{\partial} {\partial t}\vec{B} - l^2 \Box \frac{\partial} {\partial t}\vec{B} \right)- \frac{1}{c}\frac{\partial^2} {\partial t^2}\left(\vec{E} - l^2 \Box \vec{E} \right) = \frac{1}{c}\vec{j} \, .
\end{align}
Here and in the following we use that partial derivatives commute. Now we insert Eq. (\ref{eq:Max4}) which gives an equation for $\vec{E}$ alone, 
\begin{equation}
\begin{aligned}
    &     - \nabla \times\left( \nabla \times \vec{E} - l^2 \Box \nabla \times \vec{E} \right) \\
    &
    - \frac{1}{c^2}\frac{\partial^2} {\partial t^2}\left(\vec{E} - l^2 \Box \vec{E} \right) =  \frac{1}{c^2} \frac{\partial}{\partial t} \vec{j} \, .
\end{aligned}
\end{equation}
Expanding the double cross product and using Eq. (\ref{eq:Max1}) yields
\begin{equation}
\begin{aligned}
    &\Box  \left(\vec{E} - l^2 \Box \vec{E} \right) = \nabla \rho + \frac{1}{c^2} \frac{\partial}{\partial t} \vec{j} \label{eq:waveE}
\end{aligned}
\end{equation}
which is the inhomogeneous wave equation for $\vec{E}$ in the BLTP theory. A similar procedure starting with the curl of equation (\ref{eq:Max2}) yields an analogous equation for $\vec{B}$,
\begin{equation}
\begin{aligned}
    &\Box  \left(\vec{B} - l^2 \Box \vec{B} \right) = - \frac{1}{c} \nabla \times \vec{j} \, .
\end{aligned}
\label{eq:waveB}
\end{equation}

\section{Waves in vacuum} 
\label{sec:vacwaves}
\noindent
In this section we consider the source-free Maxwell equations, i.e. we assume $\rho = 0$ and $\vec{j} = 0$, and we discuss wave solutions. Because of the linearity, it suffices to study plane-harmonic waves because all other waves are linear superpositions thereof. In the next subsection we re-derive the dispersion relation for such waves in the BLTP theory, then we consider the case of standing waves.
\subsection{The dispersion relation}\noindent
We assume an electric and a magnetic field that have the form of a plane harmonic wave,
\begin{equation}
    \vec{E} (\vec{r},t) = \mathrm{Re} \big\{\vec{\mathcal{E}} \, e^{i(\vec{k} \cdot \vec{r} - \omega t)} \big\}\, , 
\label{eq:Ewave}
\end{equation}
\begin{equation}
    \vec{B} (\vec{r},t) = \mathrm{Re} \big\{\vec{\mathcal{B}} \, e^{i(\vec{k} \cdot \vec{r} - \omega t)} \big\}\, . 
\label{eq:Bwave}
\end{equation}
Here $\omega$ is real and positive and $\vec{k}$ must also be real for propagating waves. Without loss of generality, the amplitude $\vec{\mathcal{E}}$ can be chosen real. By (\ref{eq:Max4}), the amplitude $\vec{\mathcal{B}}$ is then real as well, i.e., the electric and the magnetic fields are in phase. 
Then the constitutive relations (\ref{eq:constrel1}) and (\ref{eq:constrel2}) imply
\begin{equation}
    \vec{D} (\vec{r},t) = \Bigg( 1 - l^2 \Big( \frac{\omega^2}{c^2} - k^2 \Big) \Bigg) 
    \mathrm{Re} \big\{\vec{\mathcal{E}} \, e^{i(\vec{k} \cdot \vec{r} - \omega t)} \big\} \, , 
\label{eq:Dwave}
\end{equation}
\begin{equation}
    \vec{H} (\vec{r},t) = \Bigg( 1 - l^2 \Big( \frac{\omega^2}{c^2} - k^2 \Big) \Bigg) 
    \mathrm{Re} \big\{\vec{\mathcal{B}} \, e^{i(\vec{k} \cdot \vec{r} - \omega t)} \big\} \, , 
\label{eq:Hwave}
\end{equation}
where $k = | \vec{k} |$. Note that, with $\vec{\mathcal{E}}$ and $\vec{\mathcal{B}}$ real, all four fields $\vec{E}$, $\vec{B}$, $\vec{D}$ and $\vec{H}$ are in phase. The Maxwell equations (\ref{eq:Max3}) and (\ref{eq:Max4}) require
\begin{equation}
    \vec{k} \cdot \vec{\mathcal{B}} = 0 \, ,
\label{eq:Max3wave}
\end{equation}
\begin{equation}
    \vec{k} \times \vec{\mathcal{E}} + \frac{\omega}{c} \vec{\mathcal{B}} = 0 \, , 
\label{eq:Max4wave}
\end{equation}
while the Maxwell equations (\ref{eq:Max1}) and (\ref{eq:Max2}), with $\rho = 0$ and $\vec{j} = 0$, require
\begin{equation}
    \Bigg( 1 - l^2 \Big( \frac{\omega^2}{c^2} - k^2 \Big) \Bigg) \vec{k} \cdot \vec{\mathcal{E}} = 0 \, , 
\label{eq:Max1wavevac}
\end{equation}
\begin{equation}
    \Bigg( 1 - l^2 \Big( \frac{\omega^2}{c^2} - k^2 \Big) \Bigg) \Bigg( \vec{k} \times \vec{\mathcal{B}} - \dfrac{\omega}{c} \vec{\mathcal{E}} \Bigg) = 0 \, , 
\label{eq:Max2wavevac}
\end{equation}
The wave equations (\ref{eq:waveE}) and (\ref{eq:waveB}), which follow from Maxwell's equations, take the following form with $\rho = 0$ and $\vec{j} = 0$:
\begin{equation}
    \Big( \frac{\omega^2}{c^2} - k^2 \Big) \Bigg( 1 - l^2 \Big( \frac{\omega ^2}{c^2} - k^2 \Big) \Bigg)  \vec{\mathcal{E}}  = 0 \, , 
\label{eq:waveEvac}
\end{equation}
\begin{equation}
    \Big( \frac{\omega^2}{c^2} - k^2 \Big) \Bigg( 1 - l^2 \Big( \frac{\omega ^2}{c^2} - k^2 \Big) \Bigg)  \vec{\mathcal{B}}  = 0 \, . 
\label{eq:waveBvac}
\end{equation}
We call an electromagnetic wave \emph{longitudinal} if $\vec{\mathcal{E}}$ is parallel to $\vec{k}$ and \emph{transverse} if $\vec{\mathcal{E}}$ is perpendicular to $\vec{k}$. As all equations are linear, any wave is a linear superposition of longitudinal and transverse waves.
For longitudinal waves, (\ref{eq:Max4wave}) requires
\begin{equation}
    \vec{\mathcal{B}} = 0 \label{eq:beq0}
\end{equation}
while (\ref{eq:Max1wavevac}) gives the dispersion relation 
\begin{equation}
    \omega ^2 - c^2 k^2 = \dfrac{c^2}{l^2} \, .
\label{eq:BPdisp} 
\end{equation}
The other equations give no additional information. While longitudinal waves do not exist in the standard { vacuum} Maxwell theory, as can be read from (\ref{eq:Max1wavevac}) with $l=0$, we see that in the BLTP theory they do exist and they are purely electric.
For transverse waves, (\ref{eq:Max4wave}) requires that $\vec{\mathcal{B}}$ is perpendicular to both $\vec{k}$ and $\vec{\mathcal{E}}$ and that 
\begin{equation}
    c k \big| \vec{\mathcal{E}} \big| = \omega \big| \vec{\mathcal{B}} \big| .
\label{eq:E0B0}
\end{equation}
From  (\ref{eq:Max1wavevac}) we read that either again the dispersion relation (\ref{eq:BPdisp}) or the condition 
\begin{equation}
    c k \big| \vec{\mathcal{B}} \big| = \omega \big| \vec{\mathcal{E}} \big| 
\end{equation}
has to hold. The latter equation, together with (\ref{eq:E0B0}) implies the standard Maxwell dispersion relation
\begin{equation}
    \omega ^2 - c^2 k^2 = 0 \, .
\label{eq:Maxdisp} 
\end{equation}
So there are two types of waves in the BLTP theory, with the dispersion relation given by (\ref{eq:Maxdisp}) and (\ref{eq:BPdisp}), respectively. We refer to the first as to the \emph{Maxwell modes} and to the latter as to the \emph{BLTP modes}. 
All observations that 
confirm the standard { vacuum} Maxwell theory are in agreement with the BLTP theory as well, because
the latter theory admits exactly the same modes as the former. Therefore, the BLTP theory 
can be distinguished from the Maxwell theory only if the BLTP modes are observed.
Whereas the two dispersion relations can be read directly from the wave equations (\ref{eq:waveEvac}) and (\ref{eq:waveBvac}), it was necessary to consider the Maxwell equations for verifying that $\vec{E}$ and $\vec{B}$ are in phase and that the Maxwell modes are all transverse, whereas there are longitudinal and transverse BLTP modes, with the longitudinal modes purely electric. 

{ The vacuum dispersion relation of the BLTP modes can be used to obtain the group velocity $v_{\mathrm{gr}}$, the phase velocity $\mathrm{v _{\mathrm{ph}}}$ and the refractive index $\eta$.} We find for the transverse and longitudinal BLTP modes that
\begin{equation}
 v_{\mathrm{gr}} = \frac{\text{d}\omega}{\d k} = c \, \Big(1-\frac{c^2}{l^2 \omega^2} \Big) ^{1/2} \, ,
\end{equation}
\begin{equation}
v_{\mathrm{ph}} = \frac{\omega}{k} = c \, \Big(1-\frac{c^2}{l^2 \omega^2} \Big) ^{-1/2} \,
\end{equation}
\begin{equation}
{ 
 \eta = \frac{c}{v_{\mathrm{ph}}} = \Big(1-\frac{c^2}{l^2 \omega^2} \Big) ^{1/2} \, .
}
\label{eq:vsvac}
\end{equation}
\begin{figure}
\includegraphics[width=1\columnwidth]{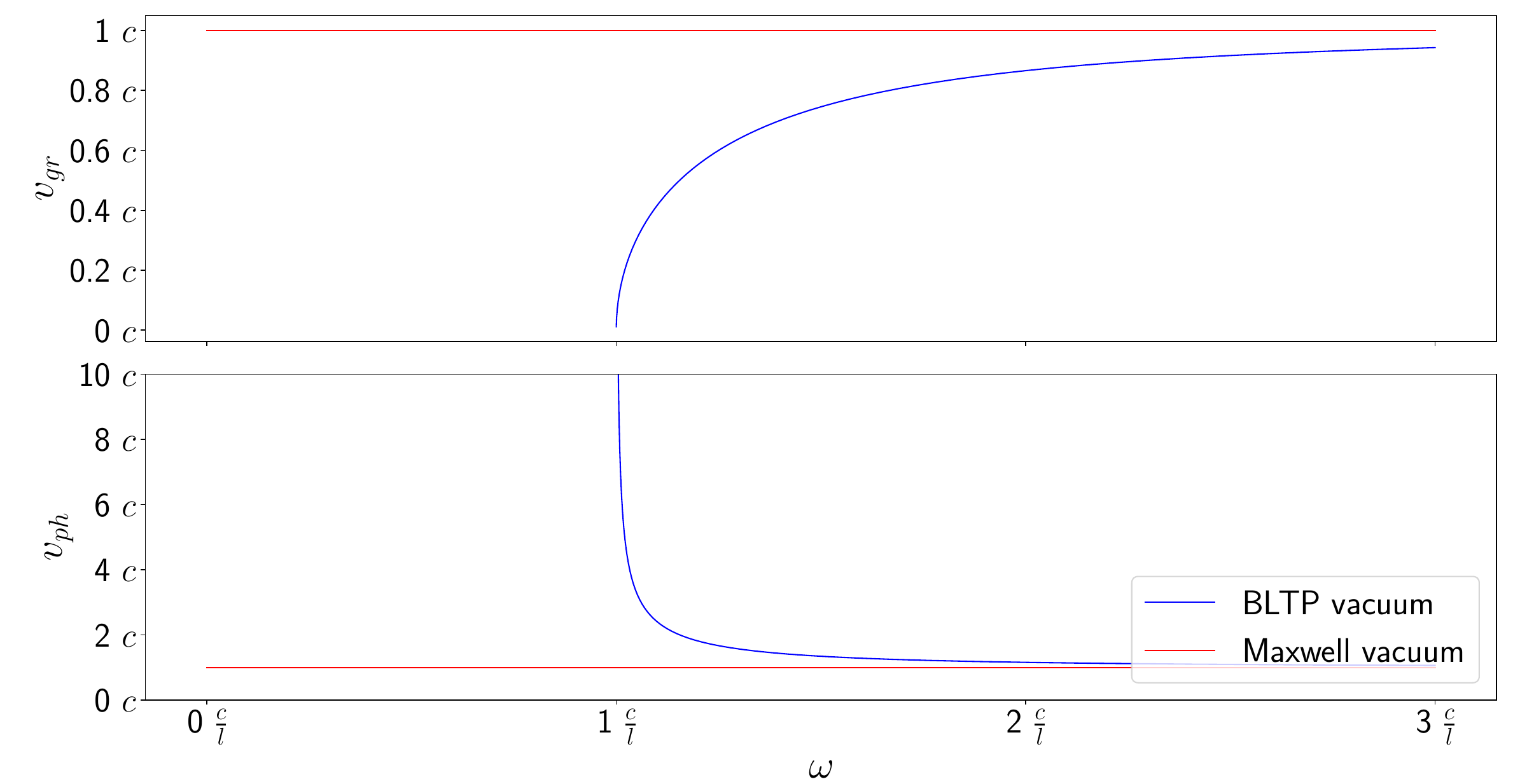}
\caption{Group and phase velocities for the (transverse and longitudinal) BLTP modes in vacuum.\label{fig:vgrwl_vac} In the case of the Maxwell modes only the transverse modes exist.}
\end{figure} \\

In Fig. \ref{fig:vgrwl_vac} we plot the phase and group velocities for the BLTP modes.

While the Maxwell modes have phase and group velocities equal to $c$, for the
BLTP modes the phase velocity is always bigger than $c$ and the group velocity is 
always smaller than $c$. The product $v_{\mathrm{ph}} v_{\mathrm{gr}}$ equals $c^2$. 

{Note that causality requires neither the phase velocity nor the 
group velocity to be limited by $c$. Only the \emph{signal velocity} must not be 
bigger than $c$. The notion of signal velocity was pioneered by Sommerfeld in 
1914. For a comprehensive discussion in English we refer to 
Sommerfeld [\onlinecite{Sommerfeld1954}], Chapter 22, and also to 
closely related work by Brillouin [\onlinecite{Brillouin1960}]. For calculating the 
signal velocity, one has to allow the frequency $\omega$ to take complex values 
and one has to view the index of refraction, $\eta$, as a complex-valued function 
on the complex $\omega$-plane. By reading the above-mentioned work by 
Sommerfeld carefully one finds that he has proven the following: Suppose that (i) 
$\eta$ has no poles in the open upper half-plane of the  complex $\omega$-plane 
and that (ii) for $| \omega | \to \infty$ the real part of the index of refraction $
\eta$ has a finite and unique limit, say $\mathrm{Re} ( \eta ) \to \eta _{\infty}$. Then the signal 
velocity equals $v_s = c / \eta _{\infty}$, i.e., it is equal to the limit of the phase
velocity for $| \omega | \to \infty$. We see that our index of refraction
(\ref{eq:vsvac}) satisfies both criteria (i) and (ii) with $\eta _{\infty} = 1$, so 
we conclude that, fortunately, the signal velocity for the transverse and longitudinal BLTP modes is equal to $c$. } 

In a quantized version, the BLTP modes correspond to a massive photon, with Compton wavelength equal to $l$. In contrast to the (de Broglie-)Proca theory, which also describes a massive photon, the BLTP theory preserves gauge invariance, see Cuzinatto et al. [\onlinecite{Cuzinatto_2017}] for a comparison of these two theories. Bopp [\onlinecite{bopp}] had suggested to identify the ``massive photon'' that occurs in his theory with the neutrino, which is of course impossible since the latter has spin 1/2. Also note that the dispersion relation (\ref{eq:BPdisp}) is identical with that for light in a cold and non-magnetized plasma with constant electron density according to the standard Maxwell theory, as was already observed by Santos [\onlinecite{SANTOS_2011}].
In this analogy, $1/l$ corresponds to $\omega _p/c$, where $\omega _p$ is the plasma frequency whose square equals the electron density up to a factor which involves only constants of nature.\\
\noindent
While the Maxwell modes exist for all frequencies, the BLTP modes exist, with a real and non-zero wave vector $\vec{k}$, only if 
\begin{equation}
    \omega ^2 > \dfrac{c^2}{l^2} \label{eq:cutoff}\, .
\end{equation}
If this inequality is violated, the dispersion relation (\ref{eq:BPdisp}) requires the wave vector to be purely imaginary, $\vec{k} = i \vec{\kappa}$ with a real vector $\vec{\kappa}$ that satisfies 
\begin{equation}
    c^2 \big| \vec{\kappa} \big| ^2 = \dfrac{c^2}{l^2}- \omega ^2  \, .
\label{eq:BPevan} 
\end{equation}
Then the equations (\ref{eq:Ewave}), (\ref{eq:Bwave}), (\ref{eq:Dwave}) and (\ref{eq:Hwave}) describe \emph{evanescent modes}, i.e., waves that do not propagate but rather fall off exponentially with an amplitude proportional to $e^{-\vec{\kappa} \cdot \vec{r}}$. 
A minimum frequency for traveling waves is well known to occur in a plasma where it is usually called the \emph{cut-off frequency}. In view of the formal analogy between the BLTP modes in vacuum and plasma waves in the standard Maxwell theory condition (\ref{eq:cutoff}) should not come as a surprise.

\subsection{Standing waves}
\noindent
Using the dispersion relation for propagating modes we specify standing waves in a one-dimensional resonator of length $L$. We assume perfectly reflecting mirrors  that are of perfectly conducting material and infinitely extended in the directions perpendicular to the wave vector. If we choose the latter to be parallel to the $x$-axis, the conditions that must be fulfilled are the Dirichlet boundary conditions, $\vec{E}(x=0,t) = \vec{E}(x=L,t) = 0$. For the transverse modes, this follows immediately from the general Maxwell equations. For the longitudinal equations it follows from the constitutive relation that links $\vec{E}$ to $\vec{D}$, because the general Maxwell equations require $\vec{D}$ to be continuous at the surface of the mirror. Therefore, in either case the electric field must be the superposition of modes
\begin{equation}
   \vec{E}{}^{(N)} (x,t) = \vec{\mathcal{E}}{}^{(N)} \, 
    \mathrm{sin} \big( k^{(N)} x \big) \, \mathrm{cos} \big( \omega ^{(N)} t \big) 
\end{equation}
where 
\begin{equation}
    k^{(N)} = \dfrac{N \pi}{L} \, , \quad N = 1,2,\dots
\end{equation}
and the frequencies $\omega ^{(N)}$ have to be distinguished for the Maxwell modes and for the BLTP modes, 
\begin{align}
    \omega_{\text{M}}^{(N)} = c\frac{N \pi}{L} \quad , \quad \omega_{\text{BLTP}}^{(N)} = c\left[\left(\frac{N \pi}{L}\right)^2+l^{-2}\right]^{1/2} \, .
\end{align}
Solving for $L$ shows that a standing wave with BLTP mode of frequency $\omega$ requires a resonator of length 
\begin{align}
 L &= \frac{1}{N\pi} \left(\frac{\omega^2}{c^2} - l^{-2}\right)^{-\frac{1}{2}} 
\end{align}
From this equation we read that the condition $\omega > c/l$ must be satisfied for having a real $L$. Unsurprisingly, this is the same condition which we already got for propagating unbounded waves. Given the current upper bound of $l < 10^{-18}\, \text{m}$ this condition requires very high frequencies which are in the gamma-ray regime: The corresponding energies $\hbar \omega$ have to satisfy
\begin{align}{
0.197 \,}
\text{TeV} 
< \hbar\frac{c}{l} 
<\hbar\omega
\, .
\end{align}
As such frequencies are much higher than lasers could produce at present, one might think of producing photons of such energies with accelerators. It is indeed true that the highest produced energies in the LHC are in the TeV regime, but it is hard to see how one could trap photons of these energies in a resonator. Therefore, testing BLTP vacuum electrodynamics in a laboratory experiment with standing waves does not seem to be feasible in the foreseeable future. In the next section we will discuss whether the situation is more promising if we consider wave propagation in a plasma, rather than in vacuum. 

\section{Waves in a plasma}

\noindent  In the preceding section we have seen that propagating BLTP modes in vacuum exist only for very high frequencies; for lower frequencies these modes are evanescent. This effect has been compared to the cut-off frequency in a plasma according to the standard Maxwell theory. The similarity between the BLTP modes in the BLTP theory and wave propagation in a plasma according to the standard Maxwell theory suggests to investigate the situation where we have an actual plasma in the BLTP theory. While we have seen that for light propagation in vacuum it is very difficult to observe the BLTP modes (if they exist), one may have some hopes that it might be easier to observe them in the presence of a plasma, because then the plasma effect could enhance the observable features of the BLTP modes. In particular, it could be possible that in a plasma of an appropriately chosen density propagating BLTP modes may exist at considerably lower frequencies than in vacuum.

Following this line of thought, we will now consider the BLTP theory in the presence of a plasma. For the latter we restrict to the simplest model, assuming a collision-less, non-magnetized and ``cold'', i.e., pressure-less, electron-ion plasma that is optically thin.

\subsection{Dispersion relation}
\noindent
In the following we derive the dispersion relation for electromagnetic waves in the BLTP theory in the presence of a plasma. We follow as closely as possible the well-known procedure for electromagnetic waves in the standard Maxwell theory in the presence of a plasma, see e.g. Jackson [\onlinecite{Jackson:100964}] or Chen [\onlinecite{chen1984introduction}].
Considering a two-fluid model for the plasma, the continuity equation and the momentum balance equation for the electron fluid read
\begin{align}
\qquad \qquad \qquad \qquad
    \frac{\partial n}{\partial t} + & \nabla \cdot (n\vec{v}) = 0 \, , 
    \label{momentum_transport}
\\
\frac{\partial \vec{v}}{\partial t} + (\vec{v}\cdot \nabla)\vec{v} & = \frac{e}{m}\left(\vec{E} +\frac{1}{c}\vec{v}\times \vec{B}\right) 
\, ,
\label{eulereq}
\end{align}
where $n$ is the number density and $\vec{v}$ is the velocity of the electron fluid in the chosen inertial system. $m (>0)$ and $e (<0)$ are the electron mass and the electron charge, respectively. 

With the constitutive relations (\ref{eq:constrel1}) and (\ref{eq:constrel2}), the inhomogeneous Maxwell equations (\ref{eq:Max1}) and (\ref{eq:Max2}) take the following form:
\begin{equation}
        \nabla \cdot \big( \vec{E} - l^2 \, \Box \vec{E} \big) =  
    e \, n + \rho_{\mathrm{ion}}
    \, , 
\label{eq:Max1plasma}
\end{equation}
\begin{equation}
        \nabla \times \big( \vec{B} - \ell ^2 \, \Box \vec{B} \big)
    - \dfrac{1}{c} \frac{\partial}{\partial t} \big(  \vec{E}- \ell ^2 \, \Box \vec{E} \big) 
= \frac{e \, n}{c} \, \vec{v} + \dfrac{1}{c} \, \vec{j} {}_{\mathrm{ion}}  \, ,
    \label{eq:Max2plasma}
\end{equation}
Here $\rho _{\mathrm{ion}}$ and $\vec{j}{}_{\mathrm{ion}}$ denote the charge density and the charge current of the ion fluid, respectively.
The homogeneous Maxwell equations (\ref{eq:Max3}) and (\ref{eq:Max4}) remain, of course, unchanged.

In the next step we linearize all equations around a homogeneous background, denoting the background with an index 0 and the perturbation with an index 1, 
\begin{equation}
n=n_0 + n_1 
\, , \quad
\vec{v}=\vec{v}_0 + \vec{v}_1 \, ,
\end{equation}
\begin{equation}
\vec{E} =\vec{E} _0 + \vec{E}_1 
\, , \quad 
\vec{B} =\vec{B}_0 + \vec{B}_1 \, ,
\end{equation}
\begin{equation}
\rho _{\mathrm{ion}} =\rho _{\mathrm{ion},0} + \rho _{\mathrm{ion},1} 
\, , \quad 
\vec{j} _{\mathrm{ion}} =\vec{j} _{\mathrm{ion},0} + \vec{j} _{\mathrm{ion},1} \, .
\end{equation}
We will later specify to the case that the perturbation is a plane harmonic wave. We assume that the electron fluid is at rest in the background, $\vec{v}_0 = 0$, with a constant density, $n_0 = \mathrm{constant}$, and that there is no background electromagnetic field,  $\vec{E} _0 = 0$ and $\vec{B}_0 = 0$ which already implies overall charge neutrality. Moreover, we assume that only the electrons but not the ions are affected by the wave, $\rho _{\mathrm{ion},1} = 0$ and $\vec{j} _{\mathrm{ion},1} = 0$, which is justified because the inertia of the ions is much bigger than that of the electrons. Then to zeroth order the inhomogeneous Maxwell equations (\ref{eq:Max1plasma}) and (\ref{eq:Max2plasma}) yield
\begin{equation} 
e \, n_0 + \rho _{\mathrm{ion}, 0} = 0 \, , \quad 
\vec{j}{}_{\mathrm{ion} , 0} = 0 \, .
\end{equation}
These two equations allow us to eliminate the charge density and the current of the ions from all equations, so these quantities will play no role in the following. The other equations are identically satisfied to zeroth order. To first order, equations (\ref{eq:Max3}) and (\ref{eq:Max4}) require
\begin{equation}
    \nabla \cdot \vec{B}_1 = 0 \, ,
\label{eq:Max3pl}
\end{equation}
\begin{equation}    
    \nabla \times \vec{E}_1 + \frac{1}{c}\frac{\partial \vec{B}_1}{\partial t} = 0 \, ,
\label{eq:Max4pl}
\end{equation}
(\ref{eq:Max1}) and (\ref{eq:Max2}) together with equations (\ref{eq:constrel1}) and (\ref{eq:constrel2}) 
require
\begin{equation}
    \nabla \cdot \Big( \vec{E}{}_1 - l^2 \Box \vec{E}{}_1 \Big) = e \, n_1 \, ,
 \label{eq:Max1pl}
\end{equation}
\begin{equation}
    \nabla \times \Big( \vec{B}{}_1 - l^2 \Box \vec{B}{}_1 \Big) 
    - \dfrac{1}{c} \, \dfrac{\partial}{\partial t } 
    \Big( \vec{E}{}_1 - l^2 \Box \vec{E}{}_1 \Big) = \dfrac{e \, n_0}{c} \, \vec{v}{}_1 \, ,
\label{eq:Max2pl}
\end{equation}
while (\ref{momentum_transport}) and (\ref{eulereq}) require
\begin{align}
&\frac{\partial n_1}{\partial t} + n_0 \nabla \cdot \vec{v}_1 = 0 \label{eq:feqomot}\\
&\frac{\partial \vec{v}_1}{\partial t} - \frac{e}{m}\vec{E}_1 = 0 \label{eq:v1-replace} \, .
\end{align}
Note that the Lorentz force produced onto the electrons by the magnetic field does not contribute to first order because, with $\vec{B}{}_0 = 0$ and $\vec{v}{}_0 = 0$, the corresponding term is of second order already.
We now assume that the electric and the magnetic field take the form of plane harmonic waves, in analogy to (\ref{eq:Ewave}) and (\ref{eq:Bwave}),
\begin{equation}
    \vec{E}{}_1 (\vec{r},t) = \mathrm{Re} \big\{\vec{\mathcal{E}} \, e^{i(\vec{k} \cdot \vec{r} - \omega t)} \big\}\, , 
\label{eq:Ewavepl}
\end{equation}
\begin{equation}
    \vec{B}{}_1 (\vec{r},t) = \mathrm{Re} \big\{\vec{\mathcal{B}} \, e^{i(\vec{k} \cdot \vec{r} - \omega t)} \big\}\, . 
\label{eq:Bwavepl}
\end{equation}
Without loss of generality, the amplitude $\vec{\mathcal{E}}$ can be chosen real. By (\ref{eq:Max4pl}), the amplitude $\vec{\mathcal{B}}$ is then real as well, i.e., the magnetic field is in phase with the electric field. By contrast, from (\ref{eq:v1-replace}) and (\ref{eq:feqomot}) we read that the velocity field $\vec{v}{}_1$ and the density $n_1$ are phase-shifted by $\pi/2$ with respect to the electric field. As a consequence, by writing 
\begin{align}
    \vec{v}{}_1 =\mathrm{Re} \big\{ i \, \vec{\mathcal{v}} \,  e^{i(\vec{k}\cdot \vec{r}-\omega t )} \big\} \, , 
\label{eq:vwavepl}
\end{align}
\begin{align}
    n_1 =\mathrm{Re} \big\{ i \, \mathcal{n} \,  e^{i(\vec{k} \cdot \vec{r}-\omega t)} \big\} \big\} \, ,
\label{eq:nwavepl}
\end{align}
we achieve that the amplitudes $\vec{\mathcal{v}}$ and $\mathcal{n}$ are real as well.

With (\ref{eq:Ewavepl}), (\ref{eq:Bwavepl}), (\ref{eq:vwavepl}) and (\ref{eq:nwavepl}) inserted into (\ref{eq:Max3pl}), (\ref{eq:Max4pl}), (\ref{eq:Max1pl}), (\ref{eq:Max2pl}), (\ref{eq:feqomot}) and  (\ref{eq:v1-replace}), we call a solution to this system of equations \emph{longitudinal} if the electric field is parallel to $\vec{k}$ and \emph{transverse} if the electric field is perpendicular to $\vec{k}$. As the system is linear, every solution is a superposition of longitudinal and transverse modes. 

We consider the longitudinal modes first, i.e., we assume that $\vec{\mathcal{E}} = \mathcal{E} \, \vec{k} / k$. Then (\ref{eq:Max4pl}), (\ref{eq:v1-replace}) and (\ref{eq:feqomot}) yield
\begin{equation}
    \vec{\mathcal{B}} = 0 \, , \quad
    \vec{\mathcal{v}} = \dfrac{e}{m \, \omega} \, \mathcal{E} \, \vec{k} /k \, , \quad
    \mathcal{n} = \dfrac{n_0 \, e \, k}{m \, \omega } \, \mathcal{E} \, , 
\label{eq:wavelong}
\end{equation}
i.e., longitudinal modes are purely electric and the velocity of the electron fluid is parallel to $\vec{k}$ as well. Equation (\ref{eq:Max3pl}) gives no additional information. Inserting (\ref{eq:wavelong}) into (\ref{eq:Max1pl}) or equivalently into (\ref{eq:Max2pl}) shows that
\begin{equation}
    \mathcal{E} - l^2 \Big( \frac{\omega ^2}{c^2} - k^2 \Big) \mathcal{E} = \dfrac{\omega _p ^2}{\omega ^2} \, \mathcal{E}
\end{equation}
where 
\begin{equation}
    \omega_p^2=\frac{n_0e^2}{m} \, .
    \label{eq:omegap}
\end{equation}
Requiring that this equation admits a non-zero solution $\mathcal{E}$ gives us the dispersion relation for longitudinal modes,
\begin{equation}
    1 - l^2 \Big( \frac{\omega ^2}{c^2} - k^2 \Big)  = \dfrac{\omega _p ^2}{\omega ^2} \, .
\label{eq:displongpl}
\end{equation}
In the standard Maxwell theory, $l = 0$, the dispersion relation reduces to $\omega = \omega _p$. This gives the phase velocity $v_{\mathrm{ph}} = \omega _p / k$ and the group velocity $v_{\mathrm{gr}} = 0$, i.e., the longitudinal modes do not describe waves that can carry a signal. They are known as \emph{plasma oscillations}, and $\omega _p$ is known as the \emph{plasma frequency}. By contrast, in the BLTP theory the longitudinal modes describe waves with  a non-zero group velocity, rather than oscillations with the fixed frequency $\omega _p$. Nonetheless, we will call $\omega _p$ as defined in (\ref{eq:omegap}) the ``plasma frequency'' also in the BLTP theory. As $\omega _p^2$ equals up to a constant factor the number density of the electron fluid, we will often refer to $\omega _p^2$ as to the \emph{plasma density}. 
With $l \neq 0$ the dispersion relation (\ref{eq:displongpl}) gives a quadratic equation for $\omega ^2$ with the solutions
\begin{align}
  \omega^2 = \frac{1}{2}c^2k^2  + \frac{c^2}{2l^2}\left(1 \pm \sqrt{(k^2l^2 +1)^2 - \frac{4 l^2 }{c^2}\omega_p^2} \right) \, . \label{eq:dispersion_longitudinal}
\end{align}
 We refer to modes where (\ref{eq:dispersion_longitudinal}) holds with the upper sign as to the \emph{longitudinal BLTP$_+$ modes} and to modes where it holds with the lower sign as to the \emph{longitudinal BLTP$_-$ modes}. A Taylor expansion of the square-root in (\ref{eq:dispersion_longitudinal}) to first order in $l^2$ shows that
\begin{equation}
    \omega ^2 = \left\{ 
    \begin{matrix} 
    \dfrac{c^2}{l^2} + c^2 k^2 - \left(1-k^2l^2+\frac{l^2}{c^2}\omega_p^2\right)\omega _p^2 + O (l^4) \, ,
    \\[0.5cm]
    \omega _p^2\left(1-\left(k^2-\frac{\omega_p^2}{c^2}\right)l^2\right) + O (l^4) \, ,
    \end{matrix}
    \right.
\end{equation}
so the frequency of the longitudinal BLTP$_+$ modes becomes infinite in the limit $l \to 0$ while the longitudinal BLTP$_-$ modes reproduce the plasma oscillations in this limit. 

\begin{figure}[h]
\includegraphics[width=\columnwidth]{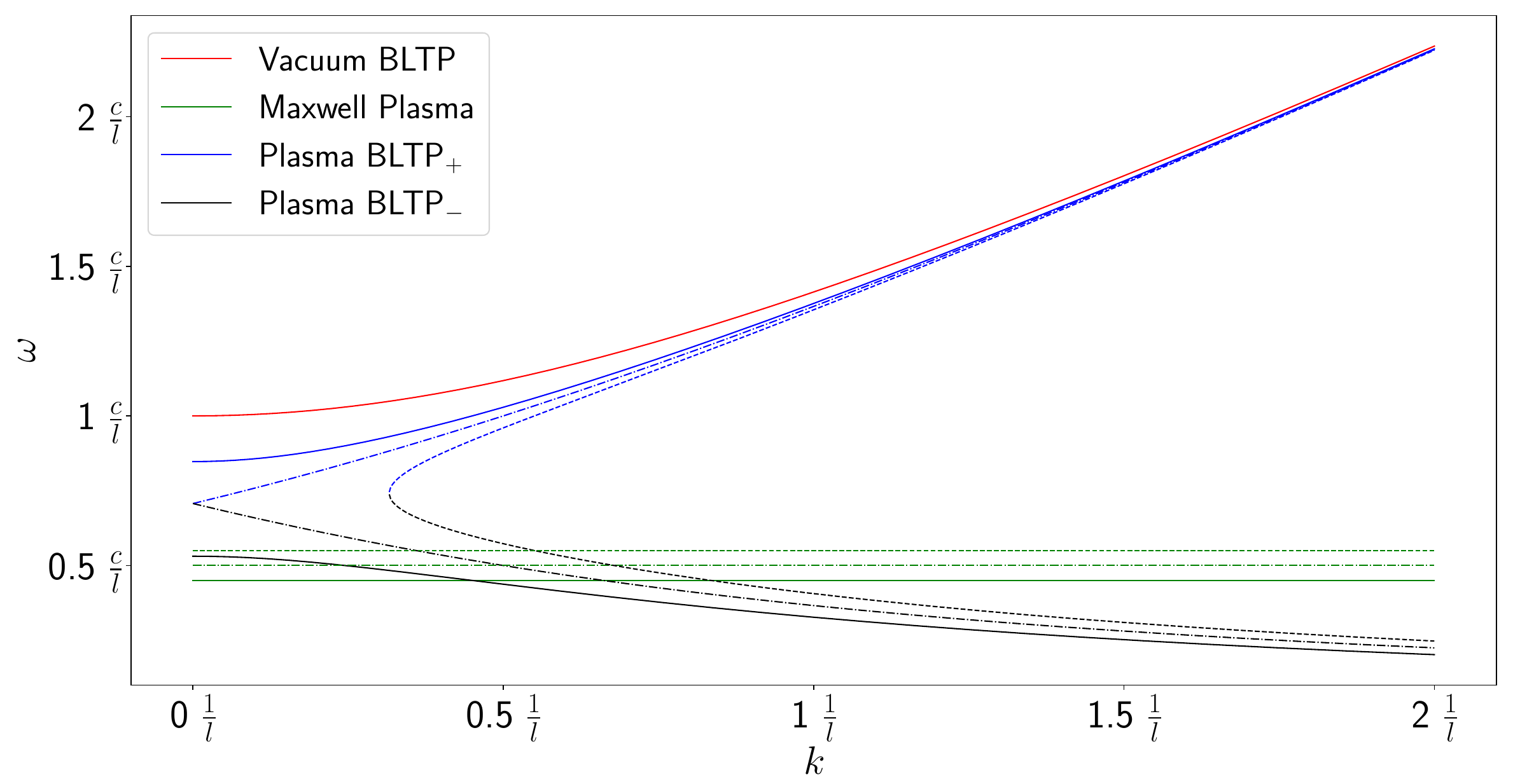}
\caption{\small Dispersion relations of the longitudinal BLTP$_+$ and BLTP$_-$ modes for a plasma frequency that is subcritical (solid blue and black lines) $\omega_p = 0.45 \frac{c}{l}$, critical (dash dotted) $\omega_p = 0.5 \frac{c}{l}$  and supercritical (dashed) $\omega_p = 0.55 \frac{c}{l}$. {For comparison the dispersion relation of the analogous Maxwell plasma using the respective plasma frequency is given in green and the vacuum BLTP is given in red.}\label{fig:displong}}
\end{figure}

\begin{figure}
\includegraphics[width=1\columnwidth]{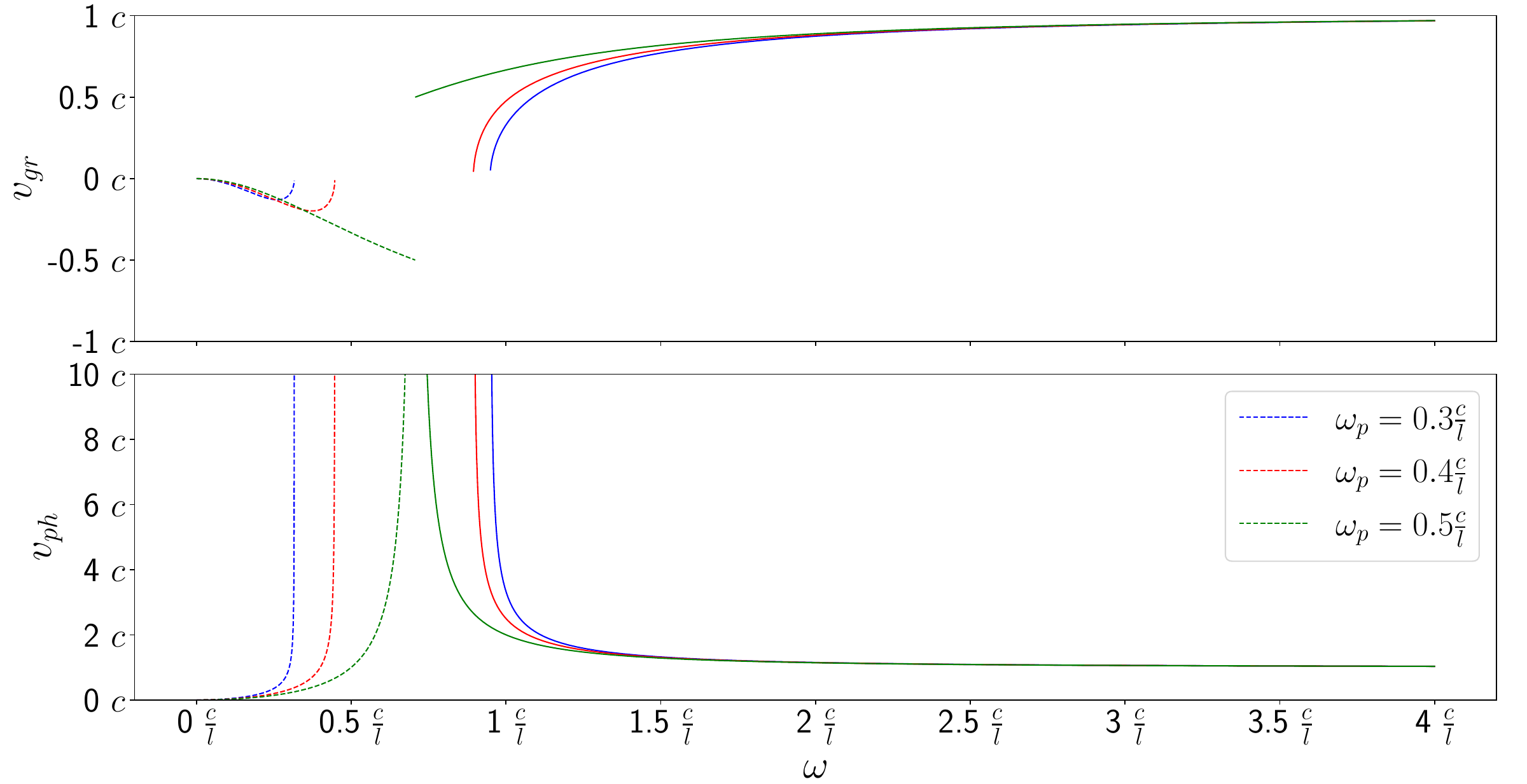}
\includegraphics[width=1\columnwidth]{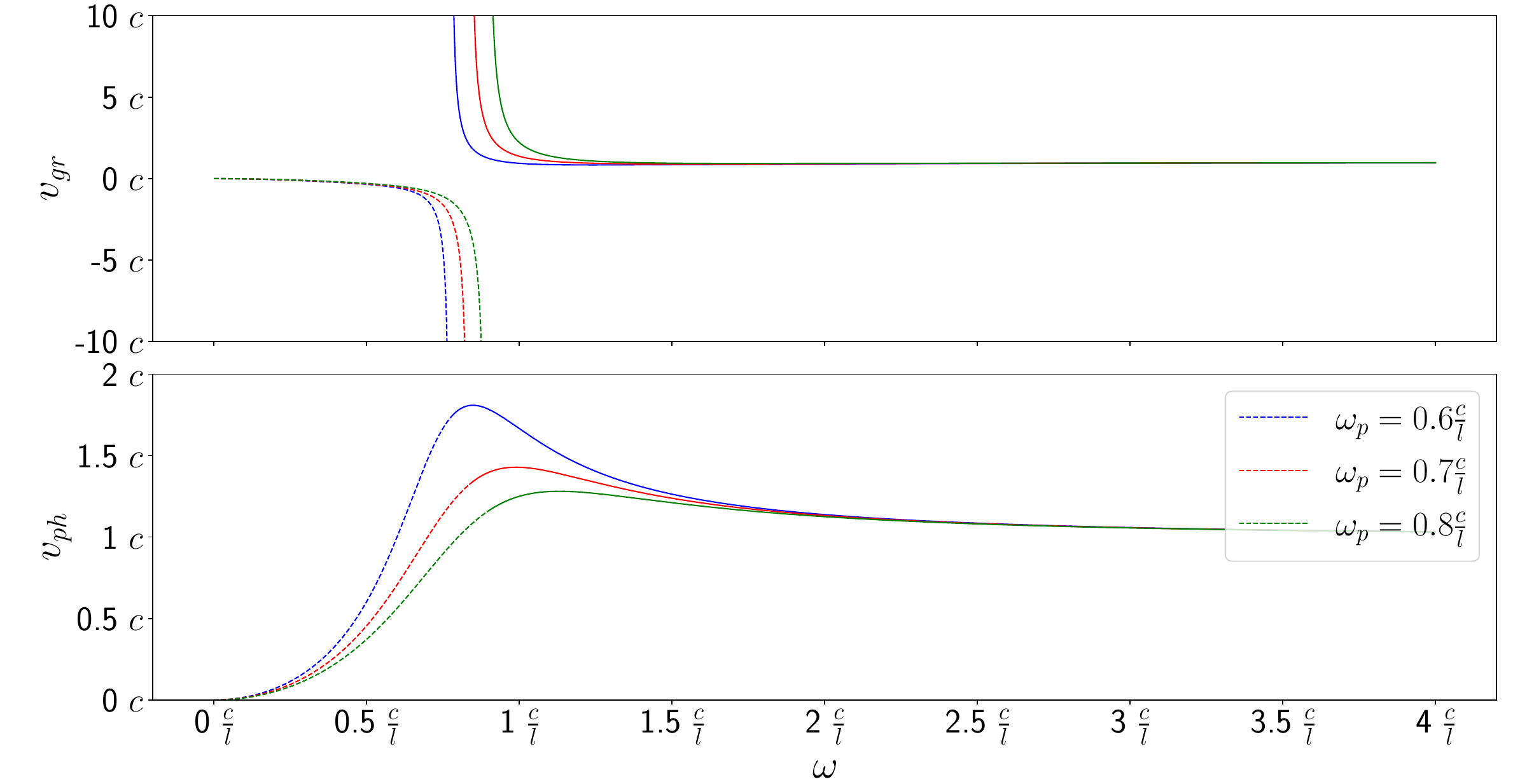}
\caption{\small Group and phase velocities for longitudinal modes in a plasma for subcritical, critical {(top panel)} and supercritical {(bottom panel)} densities. In each panel we combine the plots for the BLTP$_+$ modes (solid line) and for the  BLTP$_-$ (dashed line) in one diagram. For the BLTP$_+$ modes the group velocity is always positive and defined above a minimum frequency, for the BLTP$_-$ the group velocity is always negative and defined below a maximum frequency.  \label{fig:vgrwl}}
\end{figure}

We have to check for which frequencies propagating waves exist, i.e., for which real and positive values of $\omega ^2$ the dispersion relation is satisfied by a real and positive $k^2$. Values of $\omega ^2$ for which this is not the case describe evanescent modes. For finding the propagating modes, we first observe that the square-root in (\ref{eq:dispersion_longitudinal}) must be real, i.e.
\begin{equation}
    1+k^2 l^2 \ge \dfrac{2l}{c} \, \omega _p \, .
\label{eq:longlim}
\end{equation}
We say that the plasma density is 
\begin{align}
        &\text{subcritical if} \: \; \dfrac{2l}{c} \, \omega _p < 1 \, , 
       \nonumber
       \\[0.15cm]
        &\text{critical if} \: \; \dfrac{2l}{c} \, \omega _p = 1 \, , 
       \nonumber
       \\[0.15cm]
        &\text{supercritical if} \: \; \dfrac{2l}{c} \, \omega _p > 1 \, .
        \nonumber
\end{align}
In the supercritical case (\ref{eq:longlim}) restricts the possible real and positive values of $k^2$. If $k^2$ runs over these values, for the longitudinal BLTP$_+$ modes the frequency $\omega$ increases monotonically from a minimum value of
\begin{equation}
    \omega ^2 _{\mathrm{min}+} = \dfrac{c}{l} \, \omega _p
    \label{eq:suplonglim+}
\end{equation}
to infinity, while for the longitudinal BLTP$_-$ modes the frequency $\omega$ decreases monotonically from a maximum value of     
\begin{equation}
    \omega ^2 _{\mathrm{max}-} = \dfrac{c}{l} \, \omega _p
    \label{eq:suplonglim-}
\end{equation}
to 0. In the critical or subcritical case, (\ref{eq:longlim}) gives no restriction, i.e., $k^2$ takes all values between 0 and $\infty$. For the BLTP$_+$ modes the frequency takes its minimal value of 
\begin{equation}
\omega ^2 _{\mathrm{min}+} 
 = \frac{c^2}{2l^2}\left(1 + \sqrt{1 - \frac{4 l^2 }{c^2}\omega_p^2} \right)  
    \label{eq:longlim+}
\end{equation}
at $k = 0$ and increases monotonically to $\infty$, while for the BLTP$_-$ modes the frequency takes its maximal value of
\begin{equation}
\omega ^2 _{\mathrm{max}-} 
 = \frac{c^2}{2l^2}\left(1 - \sqrt{1 - \frac{4 l^2 }{c^2}\omega_p^2} \right)  
    \label{eq:longlim-}
\end{equation}
at $k=0$ and decreases monotonically to 0.
This is illustrated in Fig. \ref{fig:displong} where the dispersion relation (\ref{eq:dispersion_longitudinal}) is plotted. Actually, as we know that $l$ must be very small in conventional units, the supercritical case requires unrealistically high plasma densities.
As a function of $\omega$ the group and phase velocities {for longitudinal BLTP$_+$ and BLTP$_-$ modes are given by }
\begin{align}
&v_{gr} =\frac{c \, l^2 \, \omega ^2 }{l^2 \, \omega ^4- c^2 \,  \omega _p^2}
\; 
\sqrt{\omega ^4  - \dfrac{c^2 \omega ^2}{l^2} + \dfrac{c^2 \omega _p^2}{l^2}} \, ,
\label{eq:vgrlon}
 \\
&v_{ph} = 
\frac{c \, \omega ^2}{\sqrt{\omega ^4  - \dfrac{c^2 \omega ^2}{l^2} + \dfrac{c^2 \omega _p^2}{l^2}}} \, .
\end{align}
{
The respective refractive index is 
\begin{align}
&\eta = \sqrt{1  - \dfrac{c^2}{\omega ^2 l^2} + \dfrac{c^2 \omega _p^2}{\omega ^4l^2}} \, .
\end{align}
}\noindent
For the  BLTP$_-$ modes the denominator in (\ref{eq:vgrlon}) is negative.
This is obviously true in the supercritical case, where $\omega$ is restricted by (\ref{eq:suplonglim-}), but it is also true,
as a quick calculation shows, in the critical or subcritical case where $\omega$ is restricted by (\ref{eq:longlim-}). Therefore, the group velocity is negative for the longitudinal BLTP$_-$ modes. 
{ 
This means that the peak of a wave packet moves to the left while the individual plane harmonic waves it is composed of travel to the right. Given the negative group velocities one might suspect that the energy flux might be ``backwards'' as well. To check whether this is true, we calculate the Poynting vector of the modes with a negative group velocity. To that end we use the energy-momentum tensor $T^{\mu \nu}$ of electromagnetic fields according to the BLTP theory with the vacuum constitutive laws (\ref{eq:constrel1}) and (\ref{eq:constrel2}). Note that in our setting these constitutive laws are valid both for light propagation in vacuum and in a plasma, because we model the latter not by a modified constitutive law but rather by source terms. The (symmetric and gauge-invariant) energy-momentum tensor of the BLTP theory was given already by Podolsky [\onlinecite{PhysRev.62.68}], cf.  e.g. Zayats [\onlinecite{ZAYATS201411}], Eq. (8), or Gratus et al. [\onlinecite{Gratus_2015}], Eq. (128). Special care is necessary because different authors use different signature conventions.  In any case, the Poynting vector is given by the time-space components of the energy-momentum tensor; with signature $(-+++)$ it is $\vec{S} = (T^{01},T^{02},T^{03})$ (note the index positions). For a longitudinal plane-harmonic wave, i.e. (\ref{eq:Ewave}) with $\vec{\mathcal{E}}$ parallel to $\vec{k}$ and $\vec{B} = 0$, we find after a quick calculation from Podolsky's expression that     
\begin{align}\label{eq:long_poynt}
    \vec{S} = \omega \, l^2 \mathrm{sin}{}^2 \big( \vec{k}\cdot\vec{r} - \omega t \big) \mathcal{E}^2 \vec{k} \, .
\end{align}
As the scalar factor in front of $\vec{k}$ is manifestly non-negative, this demonstrates that, in the cases for which we have discovered a negative group velocity, the energy flux is \emph{not} traveling backwards. 

The group and phase velocities are displayed in fig.\ref{fig:vgrwl} for different values of the plasma frequency corresponding to the three separate cases of criticality.}

Negative group velocities have been studied in different contexts. They were first experimentally observed in epitactically grown solids by Chu and Wong [\onlinecite{Chu_solid}] and later in doped optical fibers by Gehring et. al. [\onlinecite{Gehring}]. In metamaterials they were theoretically discussed since Veselago [\onlinecite{Veselago}] and later, with the new availability of double negative materials [\onlinecite{smith}], also experimentally verified by Siddiqui et. al. [\onlinecite{Siddiqui}] and Dolling et. al. [\onlinecite{Dolling}]. { It has been stated by Dolling et. al. [\onlinecite{Dolling}] that for all combinations of signs of $v_{\mathrm{gr}}$ and $v_{\mathrm{ph}}$ in metamaterials the Poynting vector points in the ``forward'' direction, i.e., that it is proportional to $\vec{k}$ with a positive factor of proportionality.}

Obviously both the phase velocity and the group velocity go to $c$ in the limit $\omega \to \infty$. Note, however, that for the BLTP$_-$ modes $\omega$ is bounded above, i.e., this limit is defined only for the BLTP$_+$ modes. Also note that
\begin{equation}
    v_{\mathrm{ph}} \, v_{\mathrm{gr}} = 
    \dfrac{c^2 \, l^2 \, \omega ^4}{l^2 \, \omega ^4- c^2 \omega _p^2} \, .
\end{equation}

\noindent We now turn to the transverse modes, i.e., we assume that $\vec{\mathcal{E}} \cdot \vec{k} = 0$. Then 
(\ref{eq:Max4pl}), (\ref{eq:v1-replace}) and (\ref{eq:Max1pl}) yield
 \begin{equation}
    \vec{\mathcal{B}} = \dfrac{c}{\omega} \, \vec{k} \times \vec{\mathcal{E}} \, , \quad
    \vec{\mathcal{v}} = \dfrac{e}{m \, \omega} \, \vec{\mathcal{E}}  \, , \quad
     \mathcal{n} = 0 \, , 
\label{eq:wavetrans}
\end{equation}
i.e., the magnetic field and the velocity are transverse as well, with the magnetic field orthogonal to the electric field and the velocity parallel to the electric field. The density is unperturbed, i.e., the electrons oscillate without being compressed. Equations (\ref{eq:Max3pl}) and (\ref{eq:feqomot}) give no further information. Inserting (\ref{eq:wavetrans}) into (\ref{eq:Max2pl}) shows that
\begin{equation}
   \Bigg( 1- l^2 \Big( \dfrac{\omega ^2}{c^2}-k^2 \Big) \Bigg) 
   \Bigg( - \dfrac{c\, k^2}{\omega} + \dfrac{\omega }{c} \Bigg)  
   \vec{\mathcal{E}}  = \dfrac{e^2 \, n_0}{c \, \omega} \, \vec{\mathcal{E}}
   \, .
\end{equation}
Requiring that this equation admits a non-zero solution $\vec{\mathcal{E}}$ gives a 
quadratic equation for $\omega ^2 - c^2k^2$,
\begin{equation}
   \Big( \omega ^2 - c^2 \, k^2 \Big)^2
   - \dfrac{l^2}{c^2} \, \Big( \omega ^2 - c^2 \, k^2 \Big)^2
   + \dfrac{c^2}{l^2} = 0 
\end{equation}
with solution
\begin{align}
  \omega^2 = c^2k^2  + \frac{c^2}{2l^2}\left(1 \pm \sqrt{1 - \frac{4 l^2}{c^2}\omega_p^2} \right) \, .
  \label{eq:transverse_mode}
\end{align}
This is the dispersion relation for transverse modes. In analogy to the longitudinal case, we refer to modes where the upper sign holds as to the \emph{transverse BLTP$_+$ modes} and to those where the lower sign holds as to the \emph{transverse BLTP$_-$ modes}. A Taylor expansion of the square-root shows that
\begin{equation}
    \omega ^2 = \left\{ \begin{matrix} 
    \dfrac{c^2}{l^2} +c^2k^2 - \left(1+\frac{l^2}{4c^2}\omega_p^2\right)\omega _p^2 + O(l^4) \, , 
    \\[0.3cm]
    c^2k^2 + \left(1+\frac{l^2}{4c^2}\omega_p^2\right)\omega _p^2 + O(l^4) \, .
    \end{matrix}
    \right.    \label{eq:trans_taylor}
\end{equation}
We see that in the limit $l \to 0$ the transverse BLTP$_+$ modes vanish while for the BLTP$_-$ modes (\ref{eq:transverse_mode}) reduces to the well-known dispersion relation for transverse waves in a plasma according to the standard Maxwell theory, $\omega ^2 = c^2k^2+\omega _p^2$.

Next we have to discuss for which frequencies propagating transverse waves exist, i.e., for which real and positive values of $\omega$ the dispersion relation (\ref{eq:transverse_mode}) is satisfied by a real and non-zero wave-vector $\vec{k}$. We have to distinguish, again, between supercritical, critical and subcritical plasma densities. 

In the supercritical case the square-root is imaginary, i.e.
the dispersion relation has no real-valued solutions $\vec{k}$, for any real $\omega$. This means that in a plasma with supercritical  density all transverse modes are evanescent. Note that in contrast to evanescent BLTP waves in vacuum, where $k^2$ is negative, here $k^2$ is non-real, i.e., the wave-vector $\vec{k}$ has nonzero real and imaginary parts. Correspondingly, the amplitudes of these modes fall off exponentially, with an oscillating factor superimposed.  

In the critical case the dispersion relation reduces to $\omega ^2 = c^2 k^2 + c^2/l^2$, for all transverse modes, which has the same structure as the dispersion relation for transverse waves in a plasma according to the standard Maxwell theory, with $\omega _p^2$ replaced by $c^2/l^2 = 4 \omega _p^2$. This is also the dispersion relation of the BLTP vacuum modes as discussed before.

\begin{figure}[h]
\includegraphics[width=1\columnwidth]{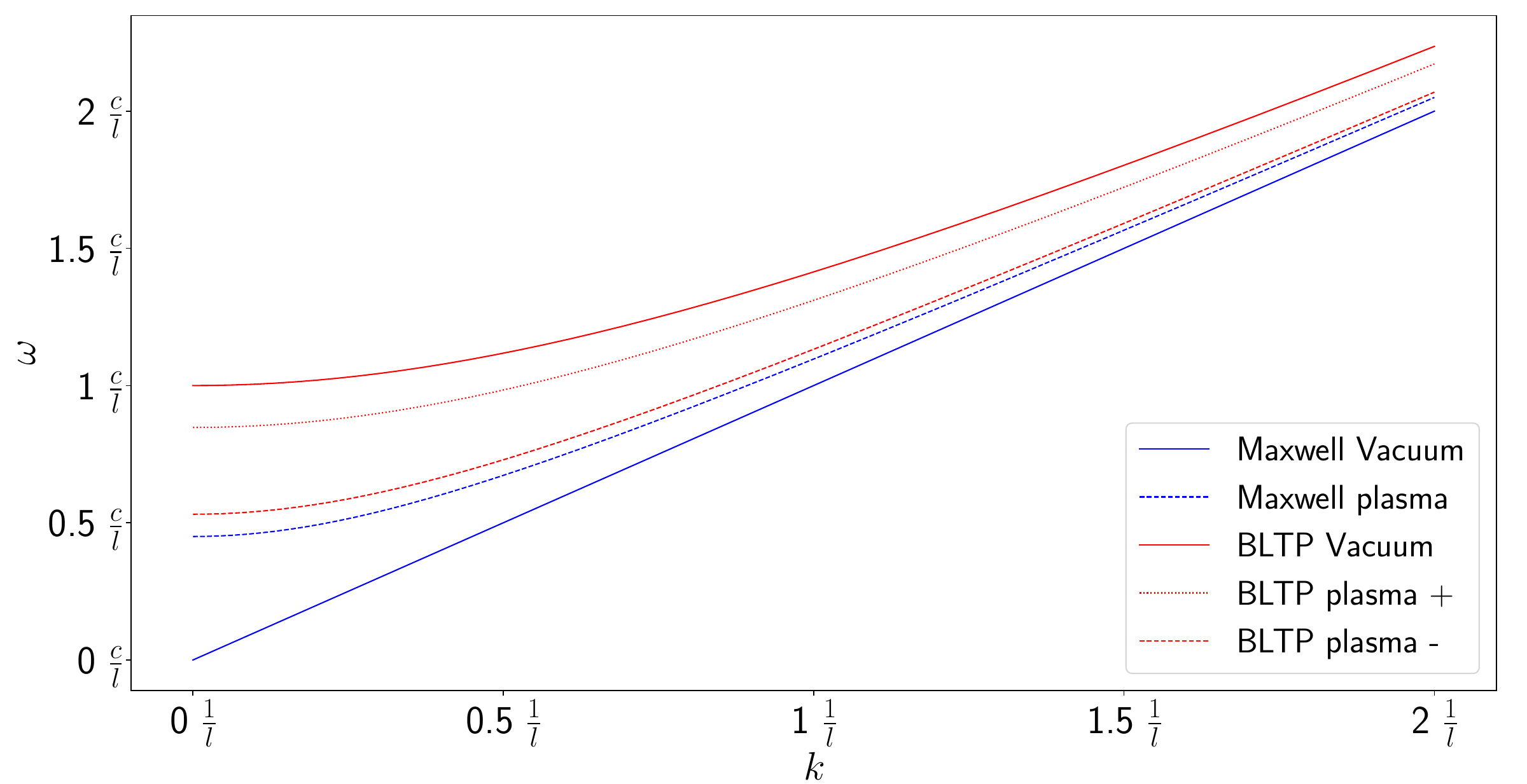}
\caption{\small Dispersion relations of transverse waves. The lines describing plasma dispersion relations use $\omega_p = 0.45 \frac{c}{l}$.\label{fig:disp_plot}}
\end{figure}

In the subcritical case the condition of $k^2$ being positive gives a lower bound for the frequency,
\begin{align}
\omega ^2  > \omega _{\mathrm{min}\pm}^2 
=
\frac{c^2}{2l^2}\left(1 \pm \sqrt{1 - \frac{4 l^2}{c^2}\omega_p^2} \right) 
\, .
\label{eq:omegaMin}
\end{align}
For real values of $\omega$ that violate this condition the modes are evanescent, with a purely imaginary wave-vector $\vec{k}$. Note that the right-hand side of  (\ref{eq:omegaMin}) is positive, for both signs. So in contrast to transverse plasma waves in the standard Maxwell theory there are now two different modes, with different cut-off frequencies both of which are lower than the vacuum BLTP cut-off frequency. The different modes are shown in fig. \ref{fig:disp_plot} for an exemplary subcritical plasma frequency of $\omega_p = 0.45\frac{c}{l}$ and a supercritical case with $\omega_p = 0.55\frac{c}{l}$.

\begin{figure}[h]
\includegraphics[width=1\columnwidth]{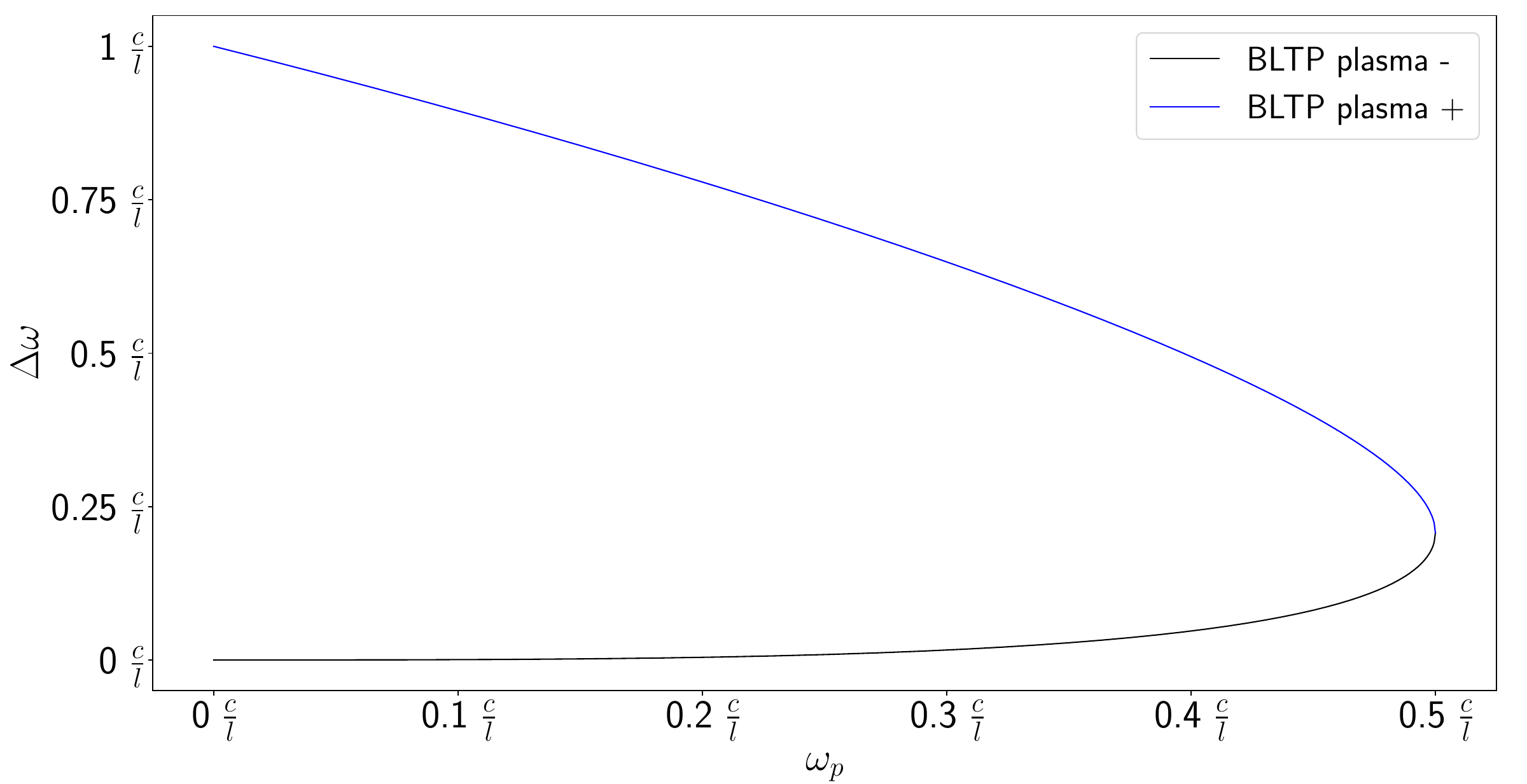}
\caption{\small Deviation from the Maxwell case of the minimum frequency for transverse modes, as a function of the plasma frequency \label{fig:disp:diff}}
\end{figure}

For the transverse BLTP$_-$ modes the minimal frequency can be arbitrarily small if the density is sufficiently small but non-zero. This makes the transverse BLTP$_-$ modes more appropriate than the transverse BLTP$_+$ modes for constraining $l$ as we will discuss later.
The deviation from the Maxwell theory is most pronounced when using the minimum frequencies as all dispersion relations go towards the vacuum Maxwell dispersion relation for infinite frequency. One can visualize the deviation in dependence of the plasma frequency by plotting the difference  
 \begin{align}
     \Delta \omega = \omega_{\mathrm{min\pm}} - \omega_p
 \end{align}
to the Maxwell minimum frequency, see fig. \ref{fig:disp:diff}.  Note that in this plot frequencies are given in units of $c/l$ which is a big number in conventional units, because of the existing upper bound on $l$. Correspondingly, a frequency shift of a fraction of $c/l$ is a big shift when expressed in SI units. Evidently given transverse modes choosing the minus sign in the subcritical regime will produce the best environment for constraining the Bopp length. We will elaborate on standing waves in the next section.

From the dispersion relation we get the following expressions for the group and phase velocities of the transverse  BLTP$_{\pm}$ modes as functions of $\omega$:
\begin{figure}[h]
\includegraphics[width=1\columnwidth]{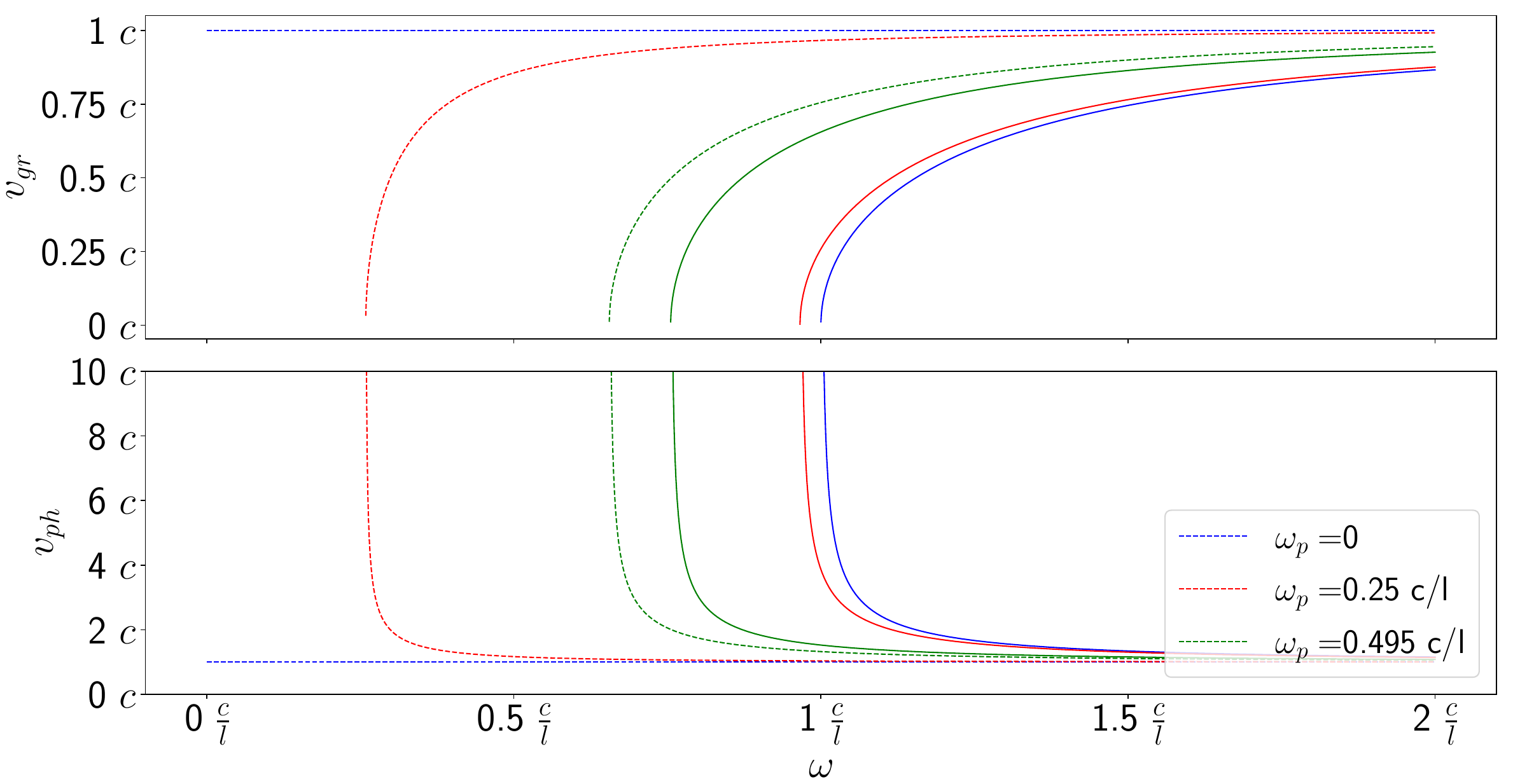}
\caption{\small Group and phase velocities for transverse modes.\label{fig:vgrwt}Dashed lines are for the BLTP$_-$ modes while solid lines are for the BLTP$_+$ modes.}
\end{figure}
{
\begin{align}
&v_{ph} = c \, \left(1  - \frac{c^2}{2l^2 \omega^2}\left(1 \pm \sqrt{1 - \frac{4 l^2}{c^2}\omega_p^2} \right)\right)^{-1/2}
\\[0.3cm]
&v_{gr} =   c \; \sqrt{1  - \frac{c^2}{2l^2 \omega^2}\left(1 \pm \sqrt{1 - \frac{4 l^2}{c^2}\omega_p^2} \right)}.
\end{align}
}
Both the phase and the group velocities are non-real for supercritical densities, which is in agreement with our earlier observation that then all modes are evanescent. Correspondingly, in Fig. \ref{fig:vgrwt} we plot these velocities only for subcritical densities. 
We see that at the minimum frequency the phase velocity diverges while the group velocity goes to zero. 
The curves for the BLTP$_-$ and BLTP$_+$ modes come together if the critical density is reached.
{ The refractive index is given by 
\begin{align}
\eta =  \frac{c}{v_{ph}} = \left(1  - \frac{c^2}{2l^2 \omega^2}\left(1 \pm \sqrt{1 - \frac{4 l^2}{c^2}\omega_p^2} \right)\right)^{1/2} \, .
\end{align}
From this expression we can calculate the signal velocity with the same method as in Sec. \ref{sec:vacwaves}. Again, we find that 
$\eta$ has no poles in the open upper half of the complex $\omega$-plane and that the real part of $\eta$ approaches 1 for 
$|\omega | \to \infty$. So also in this case the phase velocity approaches $c$ for $|\omega| \to \infty$, i.e., the signal velocity is 
equal to $c$.

Note that for the BLTP$_+$ and for the BLTP$_-$ modes also the group velocity goes to $c$ for $| \omega | \to \infty$.}
For $\omega _p \to 0$ phase and group velocities of the vacuum BLTP and Maxwell modes are, of course, recovered. Interestingly enough, for 
both phase and group velocities the differences to the Maxwell case are substantial even at low densities. 

\subsection{Standing waves}
\noindent
Similarly to the vacuum case we want to compute standing wave solutions in a plasma. We consider the idealized case of a plasma-filled resonator that consists of two infinitely extended plane parallel mirrors that are perfectly conducting and carrying no net charge. Then it follows from the general Maxwell equations that the longitudinal component of the $\vec{D}$ field and the transverse component of the $\vec{E}$ field vanish at the mirrors. Together with the constitutive law this implies that for transverse and longitudinal modes in the plasma the $\vec{E}$ field vanishes at the mirrors, so the standing waves we are looking for are of the following form:
\[
   \vec{E}{}^{(N)} (x,t) = \vec{\mathcal{E}}{}^{(N)} \, 
    \mathrm{sin} \big( k^{(N)} x \big) \, \mathrm{cos} \big( \omega ^{(N)} t \big) 
    \, , 
\]
\begin{equation}
    k^{(N)} = \dfrac{N \, \pi}{L}
    \, ,
\end{equation}
where $\omega ^{(N)}$ and $k^{(N)}$ are related by the corresponding dispersion relation. 
$L$ is the length of the resonator. 
We first consider longitudinal modes, i.e., we assume that $\mathcal{E}^{(N)}_y= \mathcal{E}^{(N)}_z = 0$. The dispersion relation for longitudinal BLTP$_{\pm}$ modes was given in (\ref{eq:dispersion_longitudinal}). We see that for the BLTP$_+$ modes the presence of the plasma is of no advantage: Standing waves exist only at exceptionally high frequencies. This is different for the BLTP$_-$ modes. Setting $k = k^{(N)}= N \pi /L$ in (\ref{eq:dispersion_longitudinal}) with the minus sign gives us the following frequency $\omega = \omega ^{(N)}$:
\[
    \omega ^2 = \dfrac{c^2 N^2 \pi ^2}{2 \, L^2} + 
    \dfrac{c^2}{2 l^2} 
    \Bigg(
    1 -\sqrt{\Big( \dfrac{l^2 N^2 \pi ^2}{L^2} + 1 \Big) ^2 -
    \dfrac{4 \, l^2}{c^2} \, \omega _p^2 }
    \Bigg)
\]
\begin{equation}\label{eq:first_order_deviation_l}
     =
    \omega _p^2 
    \Bigg( 
    1 - \Big( \dfrac{\pi ^2 N^2}{L^2} - \dfrac{\omega _p^2}{c^2} \Big) l^2 + O (l^4) 
    \Bigg) \, ,
\end{equation}
i.e., by choosing $\omega _p$ small but non-zero we get standing waves with arbitrarily small frequency. The BLTP theory predicts the existence of these longitudinal standing waves whose frequencies are close to but different from the plasma frequency.  The deviation from the plasma frequency is proportional to the term $\Big( \pi^2 N^2/L^2 - \omega _p^2/c^2 \Big)$ which can be made big by choosing a big $N$, i.e., higher harmonics, or a small resonator length $L$. 

Turning to the transverse modes we will first elaborate on the BLTP$_+$ modes. Using the lowest energy possible for these modes at the critical density of $\omega_p = 0.5 \frac{c}{l}$, the minimum frequency (eq. \ref{eq:omegaMin}) gives $\omega = \frac{c}{\sqrt{2}l}$ which differs from the vacuum case only by an irrelevant factor of $1/ \sqrt{2}$. In contrast to the longitudinal case, considering higher harmonics makes the situation worse in terms of the minimum frequency as can be seen from the Taylor expansion eq (\ref{eq:trans_taylor}). Thus the transverse BLTP$_+$ modes do not yield any advantage in view of restricting the Bopp length.

For the transverse BLTP$_-$ modes we find with the condition $k=\pi N / L$ for a standing wave 
\begin{align}\label{eq:first_order_deviation_t}
  \omega^2 =& \frac{c^2\pi^2N^2}{L^2}  + \frac{c^2}{2l^2}\left(1 - \sqrt{1 - \frac{4 l^2}{c^2}\omega_p^2} \right) 
  \nonumber
  \\ 
  =& \frac{c^2\pi^2N^2}{L^2}  + \left(1+\frac{l^2}{c^2}\omega_p^2\right)\omega_p^2 + O(l^4)\, .
\end{align}
 For increasing the deviation from the Maxwell plasma theory we have to increase the plasma frequency. Note that the term $(l^2/c^2) \omega _p^2$ is very small in comparison to 1 unless the plasma density is unrealistically high. In present laboratory experiments with optically thin plasmas the plasma density ranges from $n_0 = 10^{14}\, \text{m}^{-3}$ to $n_0 = 10^{20}\, \text{m}^{-3}$. This density is at least 16 orders of magnitude smaller than the critical density according to the BLTP theory.   
 Therefore, measuring the deviation from the Maxwell {($l=0$)} plasma theory requires to measure very small frequency differences { of the kind
 \begin{align}
     \Delta \omega = |\omega_{\mathrm{BLTP}}-\omega_{\mathrm{Maxwell}} |.
 \end{align}
Using \ref{eq:first_order_deviation_l} and \ref{eq:first_order_deviation_t}  while considering} a resonator with $L=1 \, \text{m}$ and the first mode $N=1$, plasma frequencies in the range of $5.6 \times 10^{8} \, \text{Hz}$ to $5.6 \times 10^{11} \, \text{Hz}$ yield frequency shifts of $|\Delta \omega | =1.8 \times 10^{-27}\,  \text{Hz}$ to $|\Delta \omega | =9.8 \times 10^{-19}\,  \text{Hz}$ for the longitudinal BLTP$_-$ modes and of $|\Delta \omega | =1.2 \times 10^{-28}\,  \text{Hz}$ to $|\Delta \omega | =2.4 \times 10^{-19}\,  \text{Hz}$ for the transverse BLTP$_-$ modes. { Therefore we find that the biggest shifts are produced by the longitudinal modes. Together with the much lower requirements on minimum frequencies for the BLTP$_-$ modes, we conclude that the most promising modes for discovering BLTP waves are the longitudinal BLTP$_-$ modes. However,
with the current upper bound of $l\leq 10^{-18} \, \text{m}$ even these} frequency shifts seem too small to be measurable in  a laboratory experiment in the foreseeable future.
\section{Conclusions}
\noindent
In this work we have studied traveling and standing waves according to the BLTP theory in vacuum and in an unmagnetized cold plasma. 
In vacuum, the BLTP theory admits exactly the same modes as the standard Maxwell theory but in addition a second class of modes which we called the BLTP modes. While the former correspond to the usual massless photons, the latter correspond to hypothetical massive photons. While the Maxwell modes are necessarily transverse, there are transverse and longitudinal BLTP modes. For the possible observation of the BLTP modes it is crucial that these modes are evanescent for frequencies $\omega < l/c$. With the experimental bound on the Bopp length of  $l < 10^{-18} \, \mathrm{m}$ that has been found by spectroscopic methods this implies that propagating BLTP modes can exist only for frequencies above a very high limiting frequency, corresponding to an energy of approximately {$0.197$} TeV. This makes it very difficult to observe such (traveling or standing) waves in a laboratory experiment.

In a plasma there are four types of modes which we call the longitudinal BLTP$_+$, the longitudinal BLTP$_-$, the transverse BLTP$_+$ and the transverse BLTP$_-$ modes. In the limit of vanishing plasma density they reproduce the BLTP vacuum modes while in the limit $l \to 0$ they reproduce the well-known Maxwell modes in a plasma. The (longitudinal and transverse) BLTP$_-$ modes show the most promising features in view of confronting the BLTP theory with experiments { in terms of minimum frequency}. However, even for these modes the required frequencies are still very high or the resulting frequency shift is still very low for plasma densities that can be realized in the laboratory at present. In comparison to the standard Maxwell theory in the presence of a plasma, there are several qualitatively new features. While in the standard Maxwell theory the longitudinal modes are the well-known ``plasma oscillations'' with vanishing group velocity, in the BLTP theory longitudinal modes  with non-vanishing group velocity do exist. Also, in the BLTP theory there is a critical plasma density above which transverse waves cannot exist. The most striking new feature is in the fact that the longitudinal BLTP$_-$ modes show negative group velocities for all non vanishing densities. Negative group velocities are known to exist in metamaterials and it is an interesting observation that the BLTP theory allows for them in the presence of a plasma. {A negative group velocity means that the peak of a wave packet is traveling ``backwards'' while the individual harmonic waves the wave packet is composed of travel ``forwards''. However, even in the case where the group velocity is negative, we have verified that the energy flux is propagating ``forwards'' and that the signal velocity is equal to the vacuum speed of light and also pointing ``forwards''.} 

While we have restricted here to the simple case of a cold and non-magnetized plasma, these assumptions may be generalized in future works. While the transition from a cold to a warm plasma, i.e., the introduction of a pressure, it not expected to make a major difference, interesting new features are expected if a background magnetic field is introduced. This would allow to apply the formalism to the kind of plasma that is used in fusion experiments, where for high densities the confinement is usually achieved by magnetic fields. Also, considering such a plasma on a curved spacetime would make astrophysical applications possible. For a magnetized plasma on a general-relativistic spacetime, the dispersion relation was derived on the basis of the standard Maxwell theory by Breuer and Ehlers [\onlinecite{Breuer}]. Generalizing their results to the BLTP theory would allow for astrophysical tests of this theory, e.g., by considering electromagnetic waves that travel through the Solar corona or near a magnetar.

\begin{acknowledgments}\noindent
A.Sh wishes to acknowledge the support of Deutsche Forschungsgemeinschaft, Grant Nr. 505676591, ''Generalized Maxwell theories. Theoretical  structure and experimental tests''.
\end{acknowledgments}


\providecommand{\noopsort}[1]{}\providecommand{\singleletter}[1]{#1}%

\end{document}